# Controlling the Characteristics of Nanomechanical Resonators

Anastasiia Y. Nimets[1,2], Klaus Schuenemann[1], Dmytro M. Vavriv[2],
[1]Institute of High-Frequency Technology, Technical University Hamburg-Harburg,
22 Denickestrasse, 21073Hamburg, Germany, schuenemann@tuhh.de
[2]Department of Microwave Electronics, Institute of Radio Astronomy of NAS of Ukraine,
4 Chervonopraporna Str., 61002 Kharkov, Ukraine, nimets@rian.kharkov.ua, vavriv@rian.kharkov.ua

*Abstract*—The dynamics of nanomechanical resonators driven by both low- and high-frequency signals is studied. Considering, as an example, resonators made of a doubly-clamped beam with magnetomotive driving, it is shown that three-frequency resonances arise due to the interaction of the above frequencies. Properties and characteristics of these resonances are determined analytically for linear and nonlinear modes of the resonator excitation. It is shown that in opposite to the conventional two-frequency resonance, the central frequency of these resonances, their linear dynamic range, and the critical value for the bistability onset are easily controlled by changing the frequency and the amplitude of the low-frequency driving current. The obtained results can be used when developing NMRs for sensors, filters, memory elements, and other applications.

## I. Introduction

Nanomechanical resonators (NMRs) are now considered as rather promising candidates for a number of applications, including sensors, filters and memory elements [1-3]. The resonant excitation of mechanical vibrations in a beam is the main phenomenon used when developing such devices. Typically, a two-frequency resonance with the driving frequency close to the natural frequency of the beam is exploited. However, such excitation does not provide a convenient possibility for controlling the central frequency of the resonance, its linear dynamic range, and the critical value for the bistability onset what is, however, needed for the applications. Some approaches, like for example, bending of a nanoresonator beam [4] or introducing a gate to control the beam vibrations electrostatically [5], have formerly been proposed to meet this demand.

In this paper, we demonstrate that there is a relatively simple solution to this problem that is based on adding a low-frequency (LF) spectral component to the high-frequency (HF) drive. The beam geometric nonlinearity provides an interaction of these frequencies and a formation of additional two resonances. These resonances are attractive for usage since their characteristics are easily and reversibly controlled by varying the amplitude and the frequency of the LF drive.

These ideas are illustrated by considering the dynamics of a doubly-clamped beam with magnetomotive driving. The Duffing oscillator, a generally accepted model of such resonators [6, 7], is used to model the dynamics. The mathematical correctness of the results follows from the application of the well justified averaging technique [8].

The paper is organized as follows. In Section 2, the physical and mathematical models used are presented. Section 3 deals with the conventional two-frequency resonance. In Section 4, the three-frequency resonances are studied, and methods for controlling their characteristics are described. The paper results are discussed and summarized in Section 5.

## II. Physical and Mathematical Models

For definiteness sake, we consider a popular design of NMRs [6, 7, 9, 10] made of a doubly-clamped beam as shown in Fig. 1. A current source is applied to the beam to produce a current $i(t)$. The beam is placed in a dc magnetic field $B$ what enables to excite mechanical vibrations due to a Lorentz force acting on the beam. It is assumed that the beam has small transverse dimensions compared to its length $L$.

The Euler-Bernoulli theory is generally accepted when studying the dynamics of such beams. Under the hypothesis that the spatial distribution of the beam deflection is fixed and that it coincides with that of the fundamental eigenmode of the beam, this theory yields the following Duffing equation describing the time variations of the deflection [6, 7, 9, 10]

$$\frac{d^2 y}{dt^2} + \omega_0^2 y = -2\alpha \frac{dy}{dt} - \beta y^3 + \frac{LB}{m} i(t). \qquad (1)$$

Here $t$ is the time, $\omega_0$ is the natural frequency of the fundamental eigenmode, $\alpha$ is the damping parameter related with the resonator quality factor $Q$ as $\alpha = \omega_0/(2Q)$, $m$ is the effective mass of the beam, and $\beta = (9\rho L^4)^{-1} \times (8\pi^4 E)$ is the parameter of nonlinearity, where $\rho$ is the effective density of the beam, and $E$ is Young's modulus.

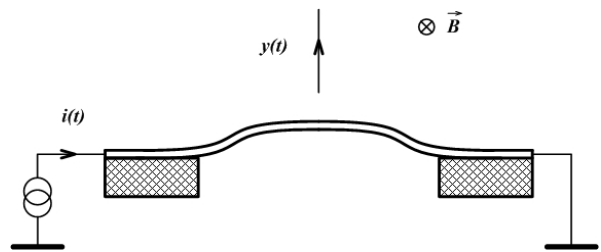

Fig. 1. Doubly-clamped beam with magnetomotive driving.

Our aim is to study effects that arise due to the interaction of LF and HF, when

$$i(t) = I_H \cos(\omega_H t) + I_L \cos(\omega_L t), \qquad (2)$$

where $I_H$ and $I_L$ are the amplitudes of the HF and LF driving currents, respectively, with corresponding frequencies of $\omega_H$ and $\omega_L$. The result of such interaction strongly depends on the relations between the characteristic frequencies (time scales) existing in the system. We consider the case when $\omega_H$ is close to the natural frequency $\omega_0$ in the sense

$$|\omega_H - \omega_0| \leq \sigma, \qquad (3)$$

where $\sigma \equiv \omega_0/Q$ is the frequency width of the resonance. As for the low frequency $\omega_L$, it is assumed to be small as compared to $\omega_H$ and $\omega_0$, but it is large with respect to $\sigma$, i.e.

$$\sigma \ll \omega_L \ll \omega_H, \omega_0. \qquad (4)$$

In order to take the low-frequency driving into account, we use the following transformation

$$y = z + C(t) \qquad (5)$$

with $C(t) = (F_L/\omega_0^2)\cos(\omega_L t), F_L = LBI_L/m$, and thus come to an equation with respect to the new variable $z$:

$$\frac{d^2z}{dt^2} + \omega_0^2 z = \frac{2F_L \alpha \omega_L}{\omega_0^2}\sin(\omega_L t) - 2\alpha \frac{dz}{dt} - \beta[z + C(t)]^3 + F_H \cos(\omega_H t). \qquad (6)$$

Considering the terms in the right-hand side as small perturbations, we can apply the averaging technique [8] to this equation. For this we introduce new variables $U(t)$ and $V(t)$,
$z = U\cos(\omega_H t) + V\sin(\omega_H t)$,
$dz/dt = -U\omega_H \sin(\omega_H t) - V\omega_H \cos(\omega_H t)$, what gives

$$\frac{dU}{dt} = -\alpha U - \Delta_0 V + \eta V[U^2 + V^2 + 4C^2(t)],$$
$$\frac{dV}{dt} = -\alpha V + \Delta_0 U - \eta U[U^2 + V^2 + 4C^2(t)] - R_H, \qquad (7)$$

where $\eta = 3\beta/(8\omega_0)$, $\Delta_0 = \omega_0 - \omega_H$, $R_H = F_H/(2\omega_H)$ and $F_H = LBI_H/m$. Thus we have nonlinear equations with periodically varying coefficients with the period of $T_v = \pi/\omega_L$ determined by the function $C^2(t)$. Similar equations have been used so far to study the stability of various physical systems and their transitions to chaos [11, 12]. In the next sections, we consider the resonances that can occur in (7).

### III. TWO-FREQUENCY RESONANCE

We start from the two-frequency resonance (3), which is usually considered when studying the dynamics of NMRs [4-7, 9, 10]. According to (3) and (4), the relaxation time ($t_{rel} \approx 1/\alpha$) of the system (7) is much larger as compared to the period of the variation $T_v$ of the coefficients in (7). Taking also into account the smallness of $|\Delta_0|$, we can apply second averaging to (7) over the period $T_v$. This leads to the equations

$$\frac{dU}{dt} = -\alpha U - \Delta_2 V + \eta V(U^2 + V^2),$$
$$\frac{dV}{dt} = -\alpha V + \Delta_2 U - \eta U(U^2 + V^2) - R_H, \qquad (8)$$

where $\Delta_2 = \omega_0 - \nu_2 - \omega_H$ with
$\nu_2 = 2\eta F_L^2/\omega_0^4 \equiv 2\pi^4 E B^2 I_L^2/(3\omega_0^5 \rho L^2 m^2)$,
which is the frequency shift of the resonant frequency due to the LF current. It is possible to show that $\nu_2$ is of the order of $\sigma$. Therefore, for this resonance, the LF current can provide just a relatively small tuning of the resonant frequency.

Considering steady states ($dU/dt = dV/dt = 0$) of (8), we arrive at the following cubic equation with respect to the intensity $W_2 = U^2 + V^2$ of the excited oscillations

$$W_2[\alpha^2 + (\omega_0 - \nu_2 - \omega_H - \eta W_2)^2] = R_H^2. \qquad (9)$$

Hence, the LF driving leads just to a frequency shift of the frequency response curve $W_2 = W_2(\omega_H)$ on the value $\nu_2$ for both linear and nonlinear modes of resonator excitation.

According to (9), the critical value of the HF driving amplitude $R_H \equiv R_H^{cr2}$, starting from which (9) has three roots (two stable ones and one unstable), is

$$R_H^{cr2} = (2\alpha/3)\bigl(2\alpha\sqrt{3}/\eta\bigr)^{1/2}. \qquad (10)$$

If $R_H > R_H^{cr2}$, the frequency response curve shows a hysteretic behaviour, as illustrated in Fig. 2. The above formula is well known from the theory of the Duffing equation, and it is used when studying the bistability of NMRs [4-7, 9, 10].

With respect to the HF driving current, formula (10) reads

$$I_H^{cr2} = 2mL\omega_0^3/(3\pi^2 BQ)\bigl(3\sqrt{3}\rho/(QE)\bigr)^{1/2}. \qquad (11)$$

The critical values (10) and (11) do not depend on the LF current. However, as will be shown below, this current has a crucial impact on these values for other resonances.

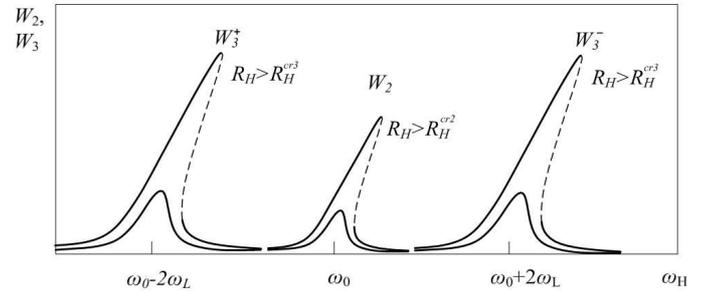

Fig. 2. Intensity of the excited oscillations vs the HF $\omega_H$ for the two-frequency ($W_2$) and the three-frequency ($W_3^\pm$) resonances.

## IV. THREE-FREQUENCY RESONANCES

In order to determine other possible resonances, we again consider (7). From the general point-of-view, we expect that this system can resonate when its characteristic frequency, which is $|\Delta_0|$, is close to the frequency of the coefficient of modulation, which is $2\omega_L$. This resonance condition assumes a pair of three frequency resonances

$$\omega_0 - \omega_H \approx \pm 2\omega_L. \quad (12)$$

Eq. (7) can be further simplified when considering these resonances. With this in mind, we introduce new variables, the amplitude $B(\tau)$ and the phase $\psi(\tau)$, as follows

$$U(t) = B(t)\cos[\pm 2\omega_L t + \psi(t)] + R_H / \Delta_0,$$
$$V(t) = B(t)\sin[\pm 2\omega_L t + \psi(t)],$$

and apply the second averaging technique over the period of the modulation coefficients. This gives the following equations

$$\frac{dB}{dt} = -\alpha B \mp \frac{\eta R_H F_L^2}{2\omega_L \omega_0^4}\sin\psi,$$
$$B\frac{d\psi}{dt} = B(\Delta_0 \mp 2\omega_L - \nu_3 - \eta B^2) \mp \frac{\eta R_H F_L^2}{2\omega_L \omega_0^4}\cos\psi, \quad (13)$$

where $\nu_3 = \eta R_H^2/(2\omega_L^2) + 2\eta F_L^2/\omega_0^2$.

From (13), we find an equation for the intensities $W_3^\pm \equiv B^2$ of the steady state oscillations for the both resonances

$$W_3^\pm[(\Delta_0 \mp 2\omega_L - \nu_3 - \eta W_3^\pm)^2 + \alpha^2] = \left(\frac{\eta R_H F_L^2}{2\omega_L \omega_0^4}\right)^2. \quad (14)$$

Nontrivial solutions of (13) and (14) can exist only due to the beam nonlinearity ($\eta \neq 0$), if the LF driving is applied ($F_L \neq 0$). This is the essential difference of the three-frequency resonances case as compared to the two-frequency resonance, where oscillations are excited by the HF drive even if $\eta = 0$ and $F_L = 0$. Nevertheless, the three frequency-resonances admit a linear mode of the NMR excitation, when the intensity of the excited oscillations is proportional to the intensity of the HF driving ($\propto R_H^2$). According to (15), this proportionality is expressed by the following formula for the resonance line

$$W_3^\pm(\omega_H) = \left(\frac{\eta R_H F_L^2}{2\omega_L \omega_0^4}\right)^2 \frac{1}{(\omega_0 - \omega_H \mp 2\omega_L - \nu_3)^2 + \alpha^2}. \quad (15)$$

It is interesting to compare (15) with an analogous formula for the resonance line of the two-frequency resonance. By using (9), we have

$$W_2(\omega_H) = R_H^2 \frac{1}{(\omega_0 - \omega_H - \nu_2)^2 + \alpha^2}. \quad (16)$$

Comparing (15) and (16), we note that these lines have the same shape with width of $1/\alpha$. However, these resonances peak at different values of the drive frequency $\omega_H$, as shown in Fig. 2. The resonance lines of the three-frequency resonances are located on the frequency scale almost symmetrically with respect to the resonance curve of the two-frequency resonance. The distance between the resonances, according to (15) and (16), is about $2\omega_L$ taking into account that the frequency shifts $\nu_2$ and $\nu_3$ are relatively small.

The lines (15) and (16) have the same intensity, if

$$\eta F_L^2 /(2\omega_L \omega_0^4) = 1, \quad (17)$$

what gives the following expression for the needed amplitude $I_L^{eq}$ of the LF current

$$I_L^{eq} = \frac{m\omega_0^2}{LB}\sqrt{\frac{2\omega_L}{\eta}} \equiv \frac{Lm\omega_0^2}{B\pi^2}\sqrt{\frac{6\rho\omega_L \omega_0}{E}}. \quad (18)$$

Provided such current amplitude is supported, the resonance lines show equal maximum intensities for all three resonances in both linear and nonlinear modes of the excitation. To estimate what values of the current are needed, we consider the NMR described in [10], which has the following parameters: $f_0 \equiv \omega_0/(2\pi)$=45.35 MHz, $B$=8T, $Q$=7770, $L$=2.25 μm, $E$=168 GPa. The beam is made of a platinum wire with the diameter of 35 nm and the density $\rho$ =21.45·$10^3$ kg/m$^3$. The characteristic frequency bandwidth of this resonator is σ/(2π) = 5.8 kHz. To satisfy the condition (4), we take the low frequency $f_L \equiv \omega_L/(2\pi)$=58 kHz. From (18), we find $I_L^{eq}$ =300 nA, what is acceptable for the applications.

The evolution of the shape of the resonance line with the transition from linear to nonlinear modes of the resonator excitation is qualitatively the same for two- and three-frequency resonances, what follows from a comparison of (9) with (14), and it is illustrated in Fig. 2. According to (14), a hysteretic loop and, consequently, bistable states arise if

$$R_H > R_H^{cr3} \equiv \frac{4\alpha\omega_L\omega_0^4}{3\eta F_L^2}\left(\frac{2\alpha\sqrt{3}}{\eta}\right)^{1/2}. \quad (19)$$

In comparison with $R_H^{cr2}$ (10), this critical value depends also on the parameters of the LF drive. Let us compare $R_H^{cr2}$ and $R_H^{cr3}$ by considering their ratio

$$\frac{R_H^{cr2}}{R_H^{cr3}} = \frac{\eta F_L^2}{2\omega_L\omega_0^4} \equiv \frac{I_L^2}{\left(I_L^{eq}\right)^2}. \quad (20)$$

Hence, the three frequency resonances offer a convenient possibility to control the bistability onset by changing the amplitude of the LF current. For example, if $I_L > I_L^{eq}$, the critical value for the bistability onset in the case of the three-frequency resonances becomes lower than that value for the two-frequency resonance. Selecting $I_L < I_L^{eq}$, it is possible to suppress the bistability onset and to increase the dynamic range of the linear mode with respect to the amplitude of the HF drive. A similar control of the bistability onset is achieved by introducing a LF modulation of the oscillator natural frequency [13], but this approach seems not very practical for NMRs.

Possible combinations of the resonance curves in a resonator with LF driving are shown on a parameter plane in Fig. 3. The coordinates of the plane are normalized amplitudes of the currents, namely $I_H/I_H^{cr2}$ and $I_L/I_L^{eq}$. In the area above the line $I_H/I_H^{cr2} = 1$, the resonance curves of the two-frequency resonance have a hysteretic loop. The curve $I_H/I_H^{cr2} = (I_L/I_L^{eq})^{-2}$ belongs to the three-frequency resonance and separates the areas with the bistable states (above the curve) and with the one-to-one resonance curves (below the curve). Hence, there are four characteristic areas on the parameter plane with different numbers (from zero to three) of resonances demonstrating bistable behaviour. By varying the amplitudes of the LF and HF currents it is possible to select different combinations of the resonances.

## V. DISCUSSION AND CONCLUSIONS

The obtained results indicate that a NMR driven by a tandem of low- and high-frequency signals demonstrates novel features as compared with harmonic driving. The interaction of such frequencies leads to the appearance of a pair of additional resonances, which are caused by a three-frequency interaction.

Although these resonances are due to the beam nonlinearity, they demonstrate a linear mode of operation with respect to the HF driving similar to that observed in the case of the conventional two-frequency resonance.

However, the parameters of the three-frequency resonances are easily and reversibly controlled by varying the amplitude and the current of the LF driving.

In particular, it is possible to increase essentially the linear dynamical range of the resonator by decreasing the current amplitude. This feature can be of interest for various sensor applications. An increase of this amplitude results in lowering the threshold for bistability with respect to the HF driving. This is obviously attractive for the development of memory elements.

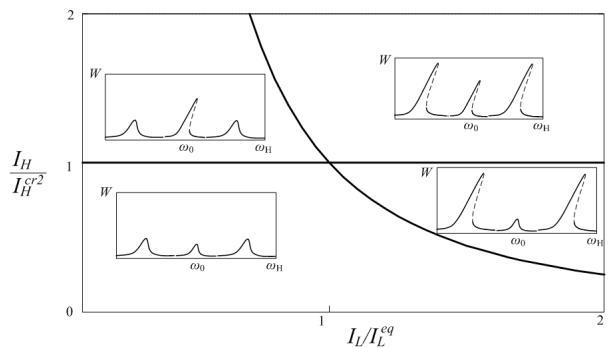

Fig. 3. Parameter plane with the normalized amplitudes of the LF- and HF currents as coordinates. The inserts show the combinations of the resonance curves of two- and three-frequency resonances for different areas on the plane.

There are totally three resonances in a NMR with the discussed tandem driving. The distance between the resonances on the frequency scale is about $2\omega_L$, where $\omega_L$ is much larger than the width of the individual resonances. Hence, the resonances are well localized in the frequency domain, and they can be used (excited) individually or collectively by selecting one, two, or three high-frequency driving with their frequencies close to $\omega_0$, $\omega_0 - 2\omega_L$, and $\omega_0 + 2\omega_L$ respectively. Practically, such single NMR is equivalent to three NMRs with harmonic driving. The central frequencies of the three-frequency resonances $\omega_0 \pm 2\omega_L$ are easily controlled in a wide (compared to the resonance bandwidth) range by varying the LF.

We have considered, as an example, a NMR made of doubly-clamped beam with the magnetomotive driving. However, due to the generality of the mathematical model used, the obtained results are applicable to other types of resonators and are not only limited to the nanosize scale.

ACKNOWLEDGMENT

A. Y. N. would like to acknowledge the DAAD for the financial support.